\documentclass[lettersize,journal]{IEEEtran}
\usepackage{amsmath,amsfonts}
\usepackage{algorithmic}
\usepackage{algorithm}
\usepackage{array}
\usepackage[caption=false,font=normalsize,labelfont=sf,textfont=sf]{subfig}
\usepackage{textcomp}
\usepackage{stfloats}
\usepackage{url}
\usepackage{multirow, ctable}
\usepackage{verbatim}
\usepackage{graphicx}
\usepackage{cite}
\usepackage{comment}
\usepackage{color,soul}
\usepackage{xcolor}
\usepackage{tablefootnote}

\begin{document}

\title{Mitigating the Effects of Au-Al Intermetallic Compounds Due to High-Temperature Processing of Surface Electrode Ion Traps}

\author{Raymond A Haltli, Eric Ou, Christopher D. Nordquist, Susan M. Clark, Melissa C. Revelle
        
\thanks{This research was supported in part by the U.S. Department of Energy, Office of Science, Office of Advanced Scientific Computing Research Quantum Testbed Program and in part by the Office of the Director of National Intelligence (ODNI), Intelligence Advanced Research Projects Activity (IARPA) under the Logical Qubits (LogiQ) program. Support is also acknowledged from the U.S. Department of Energy, Office of Science, National Quantum Information Science Research Centers, Quantum Systems Accelerator. 
Sandia National Laboratories is a multimission laboratory managed and operated by National Technology \& Engineering Solutions of Sandia, LLC, a wholly owned subsidiary of Honeywell International Inc., for the U.S. Department of Energy's National Nuclear Security Administration under contract DE-NA0003525.  This paper describes objective technical results and analysis. Any subjective views or opinions that might be expressed in the paper do not necessarily represent the views of the U.S. Department of Energy or the United States Government. 
SAND2023-14195O}}

\maketitle

\begin{abstract}

Stringent physical requirements need to be met for the high performing surface-electrode ion traps used in quantum computing, sensing, and timekeeping. 
In particular, these traps must survive a high temperature environment for vacuum chamber preparation and support high voltage rf on closely spaced electrodes.
Due to the use of gold wirebonds on aluminum pads, intermetallic growth can lead to wirebond failure via breakage or high resistance, limiting the lifetime of a trap assembly to a single multi-day bake at 200$^{\circ}$C.  Using traditional thick metal stacks to prevent intermetallic growth, however, can result in trap failure due to rf breakdown events.  Through high temperature experiments we conclude that an ideal metal stack for ion traps is Ti20nm/Pt100nm/Au250nm which allows for a bakeable time of roughly 86 days without compromising the trap voltage performance. This increase in the bakable lifetime of ion traps will remove the need to discard otherwise functional ion traps when vacuum hardware is upgraded, which will greatly benefit ion trap experiments. 

\end{abstract}

\begin{IEEEkeywords}
Wire bonding, purple plague, ion trap
\end{IEEEkeywords}

\section{Introduction}
\begin{figure}[h]
    \centering
    \includegraphics[width=0.95\columnwidth]{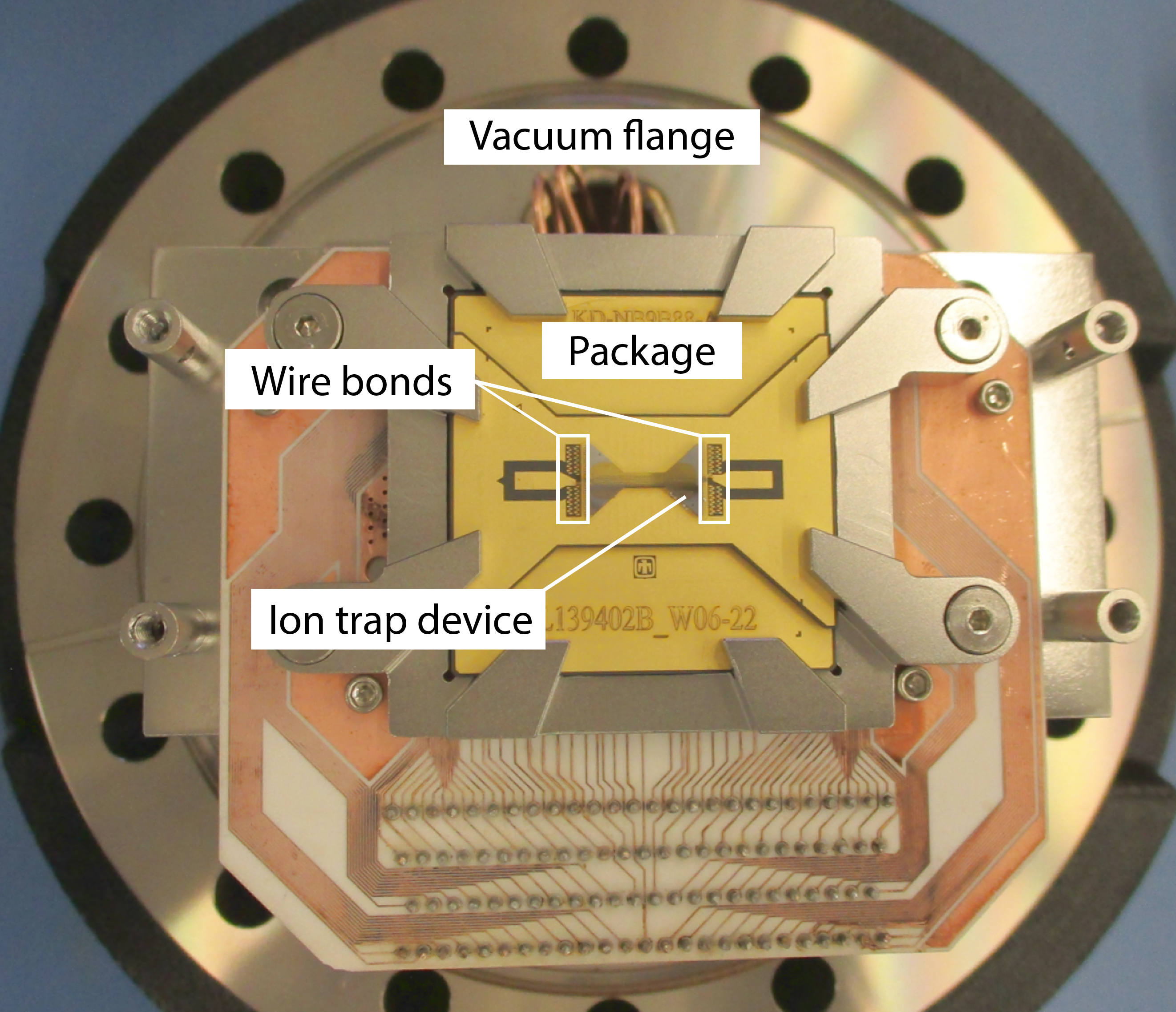}
    \caption{Photograph of a packaged ion trap mounted on the vacuum chamber device holder. Electrical signals pass through multiple electrical interfaces between the ion trap device and the electrical feed-through on the vacuum flange. To achieve the ultra high vacuum environment necessary to use ion trap devices, the vacuum chamber undergoes a high-temperature bake with the ion trap device installed. If critical faults in the electrical wiring or vacuum system components necessitate opening the chamber to troubleshoot, the ion trap device may not have the processing envelope to endure another bake without sacrificing reliability.      }
    \label{fig:iontrapchamber}
\end{figure}

\IEEEPARstart{S}{urface-electrode} ion traps~\cite{Cho2015,Tabakov2015,Maunz2016,Revelle2020,Blain2021,Moses2023} are important components to the fields of quantum computing and atomic clocks~\cite{Monroe2013,Lacroute2016,Ivory2021}.  These chip scale devices (Fig.~\ref{fig:iontrapchamber}) route voltages (dc and high-voltage rf) to electrodes which create the electric fields that confine ions tens to hundreds of $\mu$m away from the trap surface~\cite{House2008,Wesenberg2008}. 

\begin{figure}[hb]
    \centering
    \includegraphics[width=0.95\columnwidth]{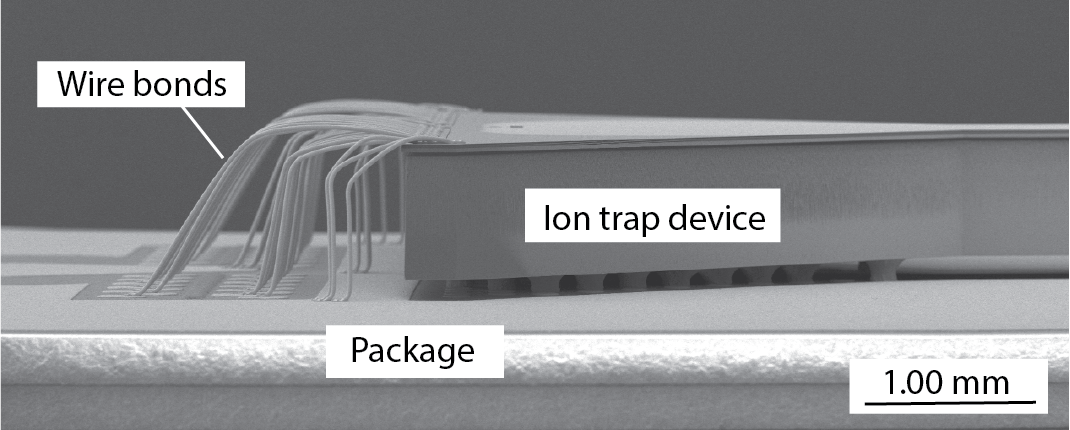}
    \caption{Scanning electron micrograph of wire bonds on an ion trap device. Wire bonds must have a low loop height to minimize interference with free space lasers necessary for trapping ions. }
    \label{fig:WirebondLoopHeight}
\end{figure}

In order to create a vacuum environment suitable for trapped-ion applications, these devices are subject to several harsh conditions.  The first is a high-temperature (200 degree C) ``bake'' for several days, which enables the ultra-high vacuum necessary to reduce background gas collisions and prevent ions from escaping the electric field confinement.  The second is the high-voltage rf field applied to the device which is necessary to confine the ions.  The high-voltage rf can lead to trap damage due to rf breakdown if the voltage is too high for the distance between the rf and ground~\cite{Wilson2022}, which is exacerbated if there is a thick metal coating applied to the top surface.  
Lastly, the devices are mounted on a package, which requires low-profile gold wire wedge bonds (Fig.~\ref{fig:WirebondLoopHeight}) to route voltages from the gold bond pads of the package to the aluminum bond pads of the device.  The combination of these conditions leads to intermetallic compound (IMC) growth~\cite{Longo1963,Selikson1964,Chen1967,Philofsky1970,Philofsky1970_2,Horsting1972,Ahmad2019} during the bake process, which if baked for too long (more than 7 days at 200$^{\circ}$C) results in trap failure due to highly resistive wire bonds or wire breaks due to Kirkendall voids~\cite{Blish2007,JI2007,Maunz2016}.  
These failure mechanisms limit the ion trap device to a single bake before potential failure, which may lead to discarding a high-performing device to fix an unrelated problem in the vacuum chamber.  
Here, we applied various metal stacks to the device bond pads with the goal of limiting failure from IMC growth without increasing the possibility of rf breakdown.

The best method for mitigating IMC growth is to avoid it by using mono-metallic interconnects, which have been shown to work well at high temperatures and in harsh environments ~\cite{Shepherd2010}.  In fact, the Au-Au system gets stronger with time and temperature~\cite{Harman2010}. Although a mono-metallic solution is preferable, it is not always practical in a research and development environment that relies heavily on common commercial and institutional processes. 
For the multi-level surface ion traps used here a mono-metallic assembly has not been available. The standard device bond pad metallization is aluminum while the standard package pad is gold.  Thus either a gold or aluminum wire bond between the 2 pads will result in an intermetallic interaction zone. 

In cases where intermetallic contacts are unavoidable, failure due to IMC growth is usually caused by high temperatures, poor welding, material impurities, and/or material defects (see Harman chapter 5~\cite{Harman2010}).
Some ways the microelectronics packaging industry has addressed this is by eliminating high-temperature processing steps, implementing automated welding, and better controlling and monitoring of impurities and defects.  
In cases where those steps cannot be taken, adding dopants such as palladium or copper when ball-bonding with gold wire has shown some promising results in reducing IMC growth~\cite{DeLucca2012,Kim2013,Xie2021}. 
In ion trapping, a Ti/Pt/Au coating is typically applied in the trapping region to prevent the growth of oxides near the ion and is thus already known to be compatible with the ion traps and voltage requirements~\cite{Blain2021}, the same coating stack has been used for years in Group III-V compound semiconductors to create ohmic contacts \cite{Seigal1996,Stareev1993}.
However, the bond pads of ion traps are not coated as the reliability of gold wedge wire bonds to Ti/Pt/Au metal on aluminum pads has not been previously reported on. 

In this paper, we report on various thicknesses of Ti/Pt/Au metal stacks to find one that is compatible with surface ion traps and that mitigates the effects of IMC growth in the wire bonds. 
Our solution is to add platinum metal to act as a diffusion barrier between the gold and the aluminum. This diffusion barrier can slow down or eliminate the harmful effects of the Au-Al IMC. 
In some systems, this solution can be simpler to implement than other solutions---such as adding dopants to the gold wire---and is suited for systems that would benefit from a gold coating. 
Our first investigation showed thick platinum and gold metal layers best mitigated the effects of IMC growth. 
However, thick metal stacks pose an issue for ion traps because they significantly reduce the electrodes gaps which increases the chance of electrical arcing due to the high voltage rf that is applied to the trap ~\cite{Wilson2022}.
In our trap designs, the electrodes overhang a lower metal layer with vertical gaps as small as $2~\mu$m \cite{Revelle2020}. 
If a 1~$\mu$m thick gold coating is applied to these traps, the vertical distance is decreased significantly, which will reduce the maximum applicable voltage. 

For ion traps a particular solution to the IMC growth in wire bonds is to use a fairly thin gold coating stack Ti20nm/Pt100nm/Au250nm in both the bond pad region and the area where the ions are trapped. This metal stack enables an ion trap to undergo multiple multi-day, high-temperature bakes without suffering the negative effects of IMC growth. Although there is still some IMC growth with this solution, it does not lead to a failure in either increased resistance or mechanical failure of the wire bonds. This solution could potentially be used for other applications that experience harsh environments including automotive, aerospace, rf systems, drilling and mining.

\section{Experimental Setup and Procedure} 
 
This section covers the experimental setup including materials choices, test device fabrication and procedures used to validate the IMC mitigation in multi-level surface electrode ion traps. 

The electrical interconnects on ion trap devices are formed using 1 mil round gold wire thermosonically wedge bonded by an automated wire bonder to connect the bond pads on the die to the bond pads on the package (Fig.~\ref{fig:WirebondLoopHeight}).
Wedge wire bonding has lower loop heights, finer pitch, and higher reliability than ball bonding ~\cite{JI2007}. Lower loop heights are important to avoid the clipping of laser light delivered across the trap surface for ion manipulation. The finer pitch of wedge bonding enables more pads on the device so a higher number of electrodes can be individually controlled in more complex ion trapping systems.
\begin{figure}[h]
    \centering
    \includegraphics[width=0.95\columnwidth]{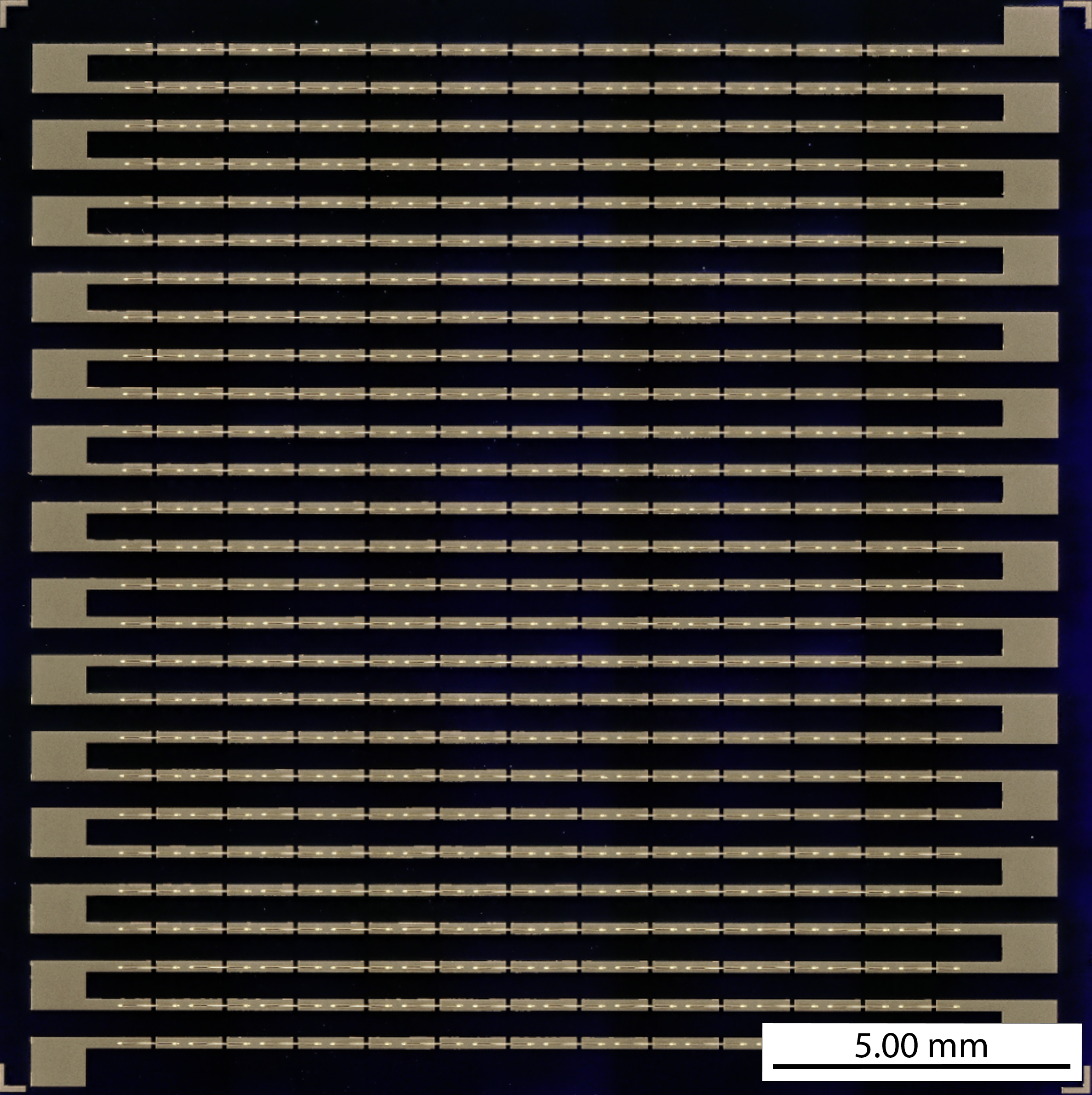}
    \caption{Optical image of the microfabricated test structure used for wire bond aging experiments. The metal pads on the chip are connected serially with wire bonds and the large bends at the ends of each row allow for probe access during electrical testing.   }
    \label{fig:TestStructure}
\end{figure}

\begin{figure}[h]
    \centering
    \includegraphics[width=0.95\columnwidth]{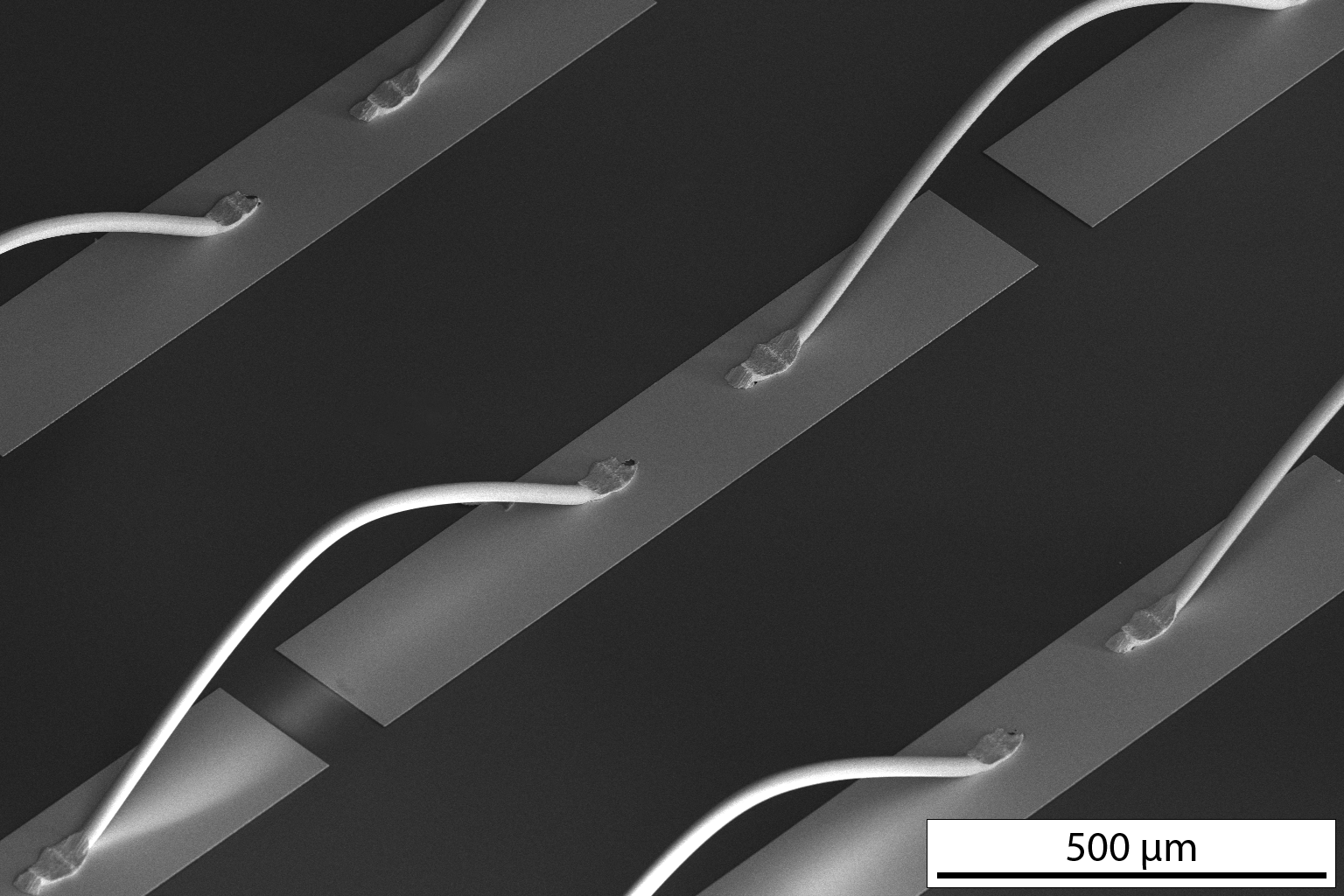}
    \caption{Scanning electronic micrograph of wire bond test structure.}
    \label{fig:TestStructure_perspective}
\end{figure}

Ion traps typically experience wire bond failure at the heel of the bond on the device, which is inherently the weakest point. The strength of the heel can be optimized by changing the wire bond parameters of ultrasonic power, force, and time. The weakness of the heel is exacerbated by the Au-Al IMC growth during the high-temperature bakes necessary to prepare the UHV conditions for ion trapping. IMC growth affects both the mechanical strength and the electrical resistance of the bond. To mitigate this failure mechanism test devices were fabricated to further investigate the cause and possible solutions.

The test device consists of a segmented metal serpentine with 27 linear test sections, each accommodating 12 wire bonds in each section, for a total of 324 bonds (see Fig.~\ref{fig:TestStructure}. and Fig.~\ref{fig:TestStructure_perspective}).
Large contact pads at each bend of the serpentine facilitate electrical testing. The test devices are fabricated on 200~mm silicon wafers using standard microelectronic fabrication techniques. First, a passivating silicon dioxide film is grown on the wafers using thermal oxidation. The base metal layers consisting of a 50~nm titanium nitride adhesion layer and 1.2~$\mu$m thick Al-0.5wt\%Cu film are deposited using physical vapor deposition. These blanket films are then patterned using photolithography and reactive ion etching (RIE) in $CF_4$ plasma. Additional metal films are deposited onto the base metal layer using a metal lift-off process. The metal stacks of interest are deposited using E-beam evaporation. Titanium is deposited as an adhesion layer, then platinum is deposited to slow or stop the gold from diffusing, and lastly the gold is deposited. See Table~\ref{tab:Failure_values} for the list of metal stacks reported on in this paper.

A combination of test methods is used to evaluate wire bond integrity. This includes pull tests, contact resistance, scanning electron micrograph (SEM) imaging, optical imaging, focused ion beam (FIB) cross-section imaging, and accelerated aging tests. 
The pull test is the most common method for testing wire bonds. It is performed by precision automated equipment to measure the breaking force of a bonded wire. In wedge wire bonding there are multiple wire break failure modes, which includes mid-span, heel, wedge, and crater. During the pull test, the sample is securely mounted to the tester stage, a small hook is aligned mid-span of the wire, and the wire is pulled up until the wire breaks. Throughout the pull test the force is recorded in grams of force (gf). Although a mid-span break generally indicates the best bond parameters, most of the time the wire breaks at the heel of the bond which is the weakest point for bonds with good welds. If the weld is bad, then the wire bond can lift off the pad. During the lifetime of a wire bonded device the failure mode can change. For example, as IMC grows, large Kirkendall voids can form and the bond itself can lift off the pad leaving a crater. Ion traps have typically suffered from heel breaks if they were exposed to excessively long bakes or multiple bakes at high temperatures.

The contact resistance of each wire bond in this study is initially in the milliohm range and therefore best measured using a 4-point resistance measurement. A Keithley 2000 multimeter with built in 4-point measurement capability is connected to a probe station to perform these measurements. This method naturally excludes the resistance of the test setup, including the meter and test leads, and precisely measures the resistance of the circuit under test.  The circuit on our test device consists of all the bond pads and wires between any two large contact pads. The resistance contribution of one bond pad and half of each wire is treated as the initial bond resistance of a weld. In general, the bond resistance increases over time but at different rates for the different metal stacks.

Thermal aging and environmental tests in conjunction with the other methods are useful for reliability testing. Thermal aging accelerates the possible failure modes; in general, this means higher temperature or faster temperature cycling. There are a few temperatures that are relevant for ion trapping experiments and accelerated aging tests. The temperature of most interest is 200$^{\circ}$C, which is the standard temperature at which the vacuum system is baked for multiple days to prepare the ultra-high vacuum environment in the range of 10e-11 Torr. Therefore, initial samples were tested for many days at 200$^{\circ}$C. Another relevant temperature range is 300$^{\circ}$C to 350$^{\circ}$C which may be necessary to attach a vacuum sealed lid to the trap package. Higher temperatures are only needed for a short time to melt the solder, which could take from minutes to a few hours depending on thermal ramp rates for the required process. These higher temperatures are also used for accelerated aging studies. 

\begin{figure}[ht]
    \centering
    \subfloat[]{\includegraphics[width=0.95\columnwidth]{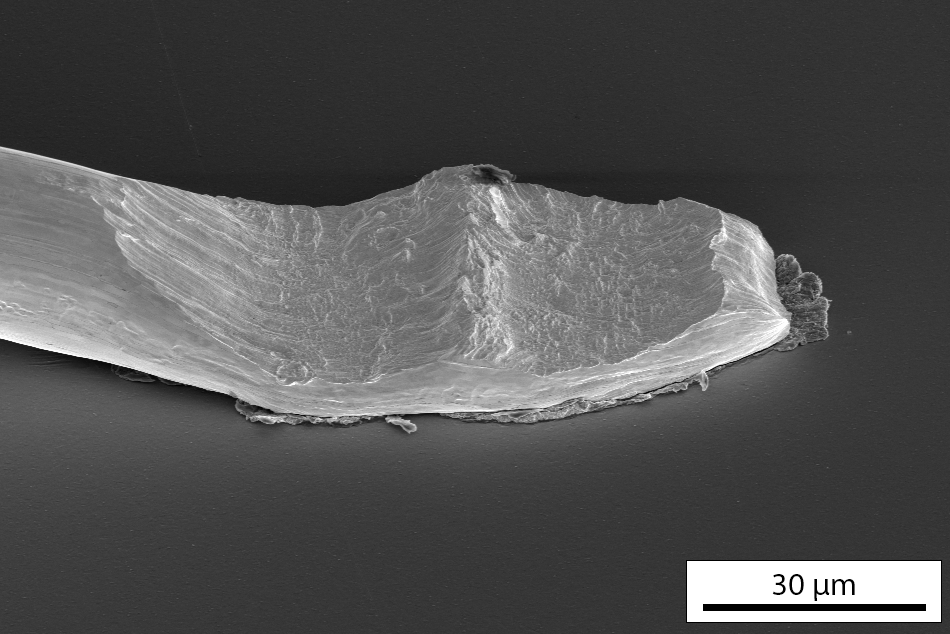}
    \label{fig:SEM_no_bake_perspective}}
    \hfil
    \subfloat[]{\includegraphics[width=0.95\columnwidth]{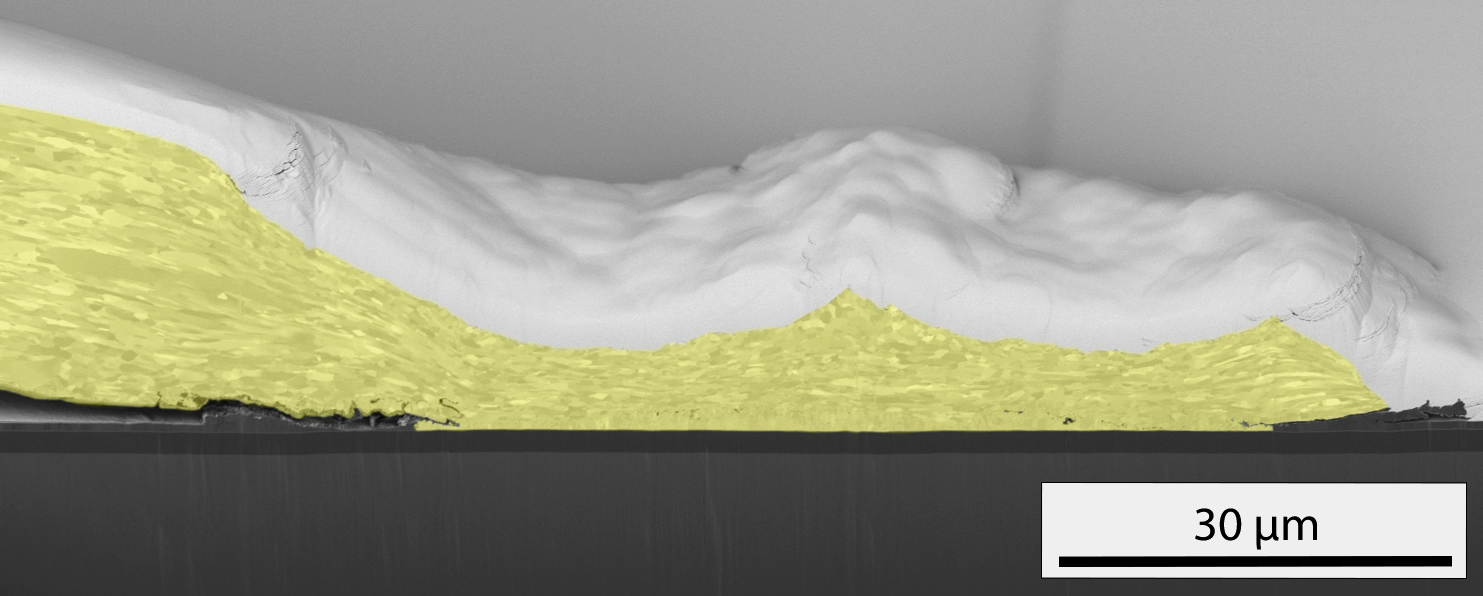}
    \label{fig:SEM_no_bake_section}}
    \caption{Scanning electron micrographs (SEMs) of a gold wire bond on bare aluminum top metal as-bonded. (a) Perspective view and (b) cross-section view after plasma focused ion beam sectioning. The false coloring is used to highlight the gold wire bond, differentiating it from tungsten and material redeposited during FIB processing.   }
\end{figure}

\begin{figure}[ht]
    \centering
    \subfloat[]{\includegraphics[width=0.95\columnwidth]{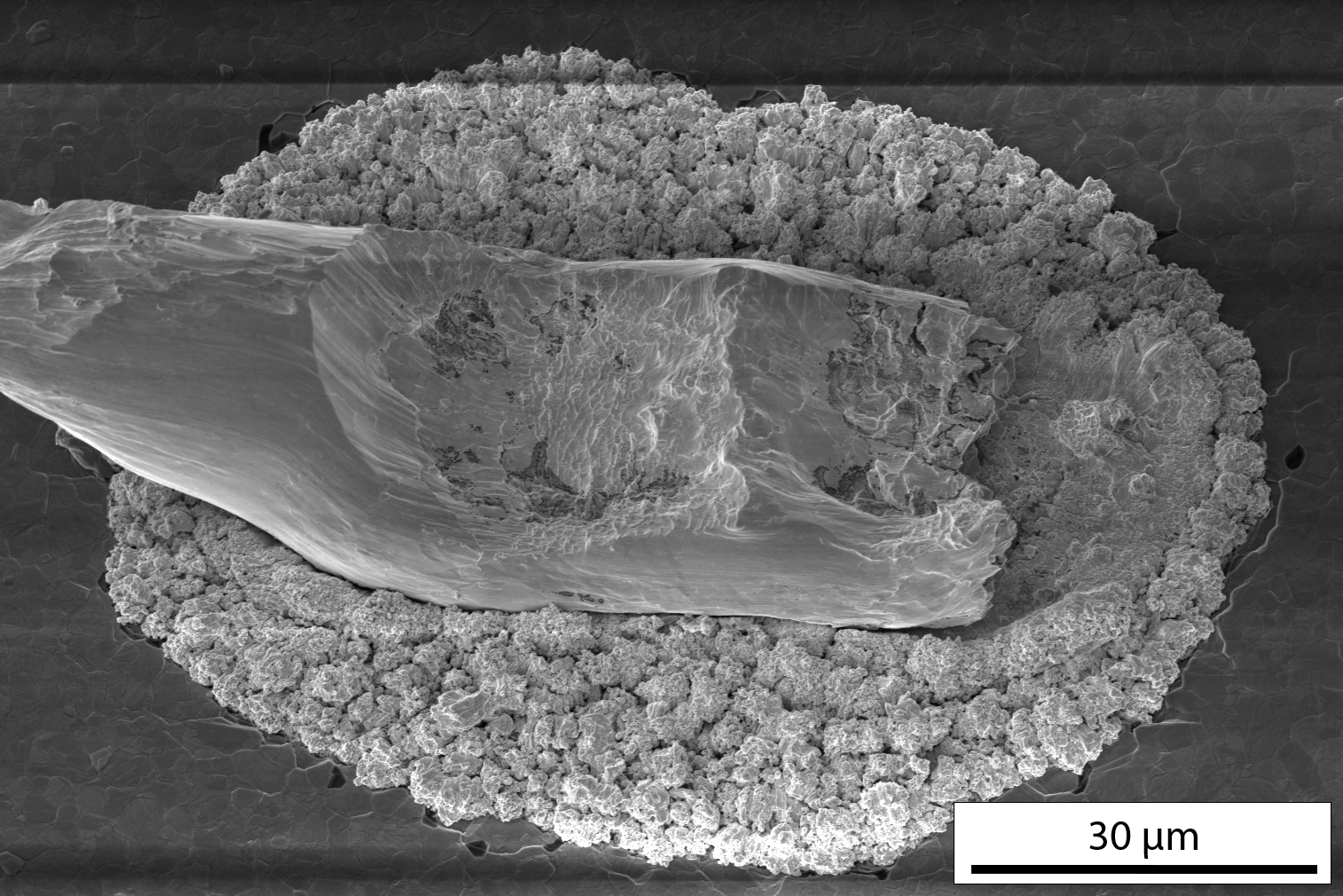}
    \label{fig:SEM_29at200_perspective}}
    \hfil
    \subfloat[]{\includegraphics[width=0.95\columnwidth]{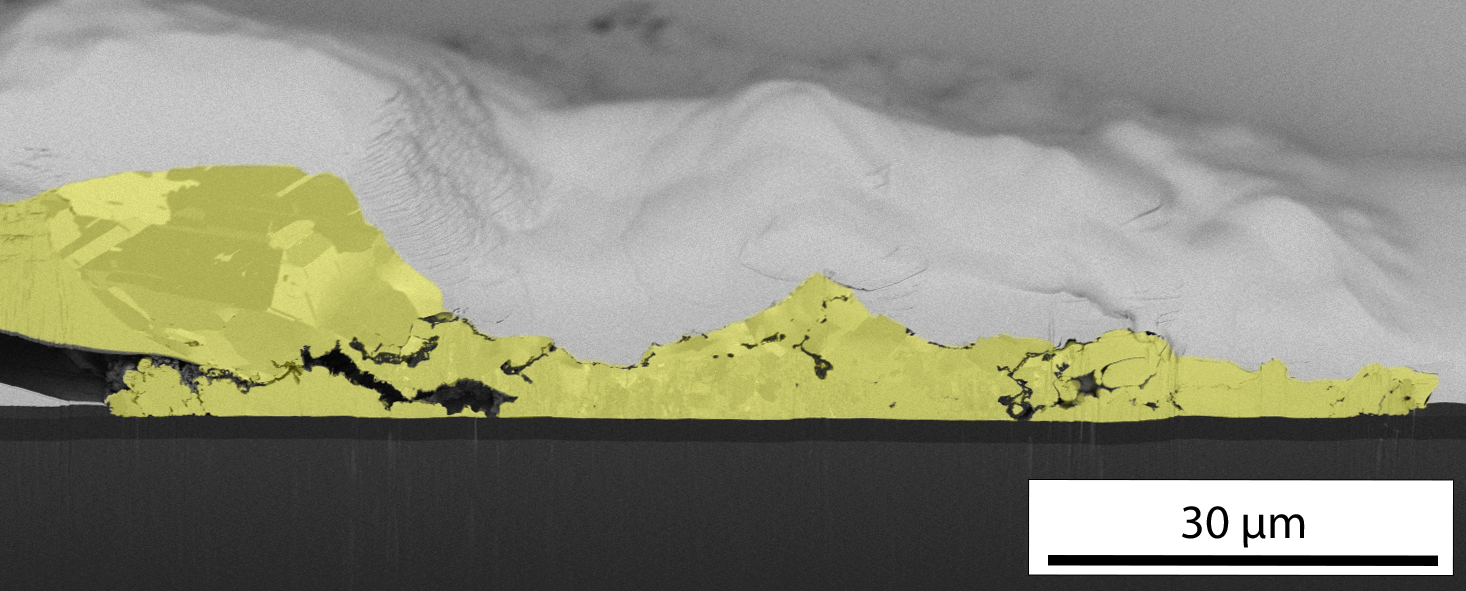}
    \label{fig:SEM_29at200_section}}
    \caption{Scanning electron micrographs (SEMs) of a gold wire bond on bare aluminum top metal after baking at 200$^{\circ}$C for 29 days. (a) Perspective view  and (b) cross-section view after focused ion beam sectioning. The false coloring is used to highlight the gold wire bond, differentiating it from tungsten and material redeposited during FIB processing. }
    \label{fig:PP_Wirebond}
\end{figure}

For each test device optical pictures of one bond on each sample are taken at regular intervals to monitor the growth of IMC around the wire bond.  Resistance measurements and pull tests are also taken at these same intervals after high-temperature bakes. SEM inspection of FIB cross-sections are performed on some samples before and after the full bake cycle, Kirkendall voids and IMC growth are observed (see Figs~\ref{fig:SEM_no_bake_section} and~\ref{fig:SEM_29at200_section}). These SEM images are also useful to note the thickness of the wire bond heel and further illustrates how it is generally the weakest point in the wire bond.

It is interesting to note that gold wire bonding to aluminum creates some initial IMC that is necessary to form the weld between the two metals. There is at least one report on etching away the bond pad metal to show IMC under the bond that gives insight into the extent of the weld and IMC growth ~\cite{Schneider2009} but this was done on ball bonds and is beyond the scope of this work. 
However, it seems that with thick enough gold deposited on a bond pad that the weld would not disturb the underlying diffusion barrier layer that protects the aluminum from Au-Al IMC growth.

\section{Results}

\begin{figure*}[!t]
    \centering
    \includegraphics[width = 0.95\textwidth]{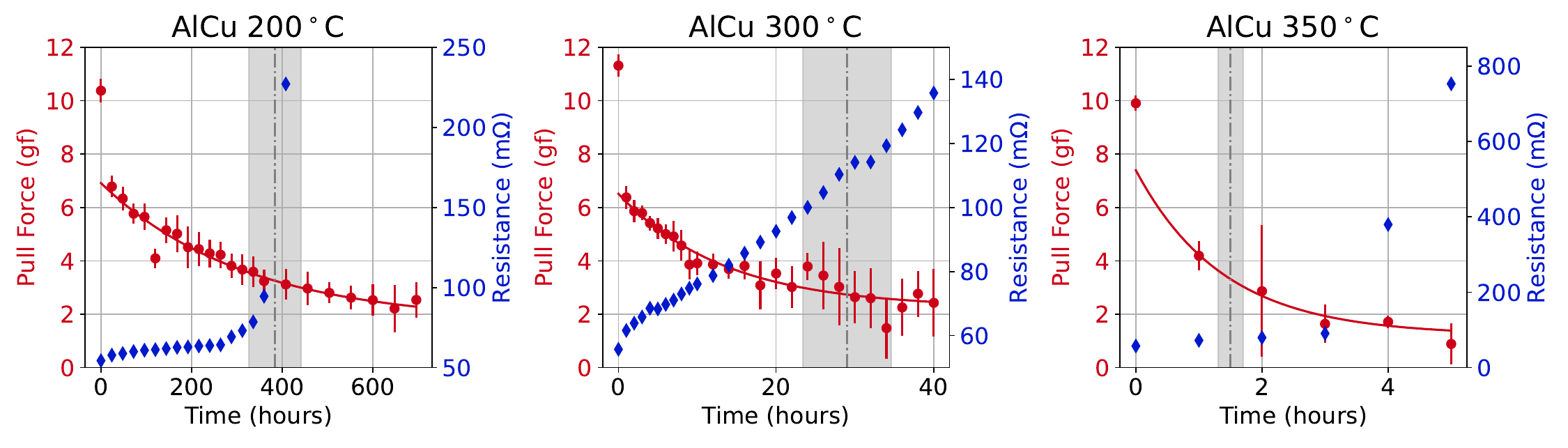}
    \includegraphics[width=0.95\textwidth]{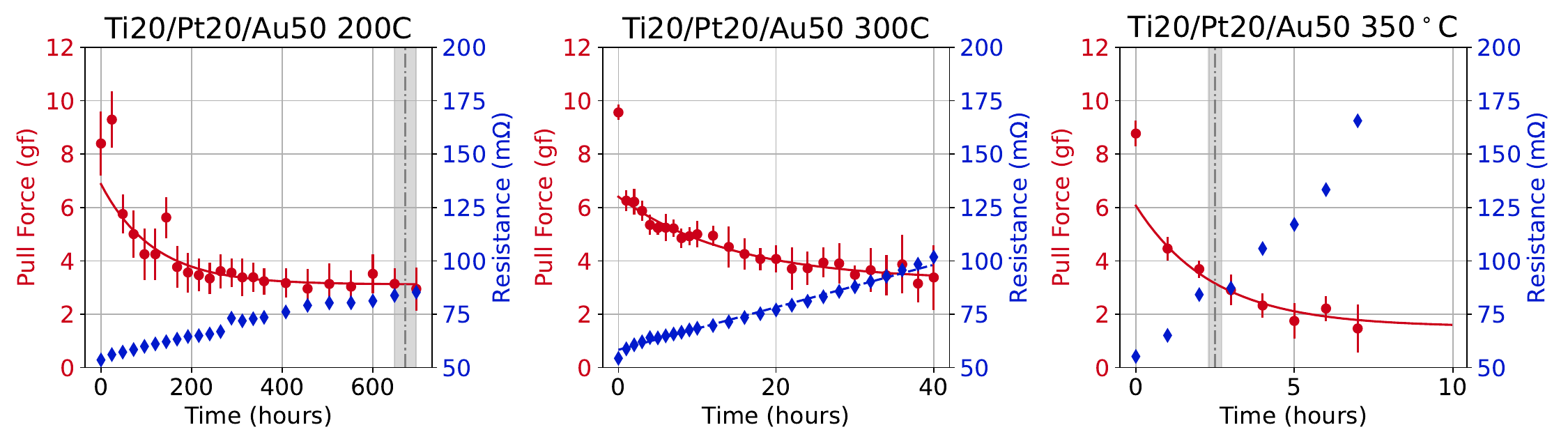}
    \caption{Measured pull force to break the wire bond and the resistance over time for a wedge bonded gold wire on (top row) {1.2\,$\mu$m} AlCu  and (bottom row) the Ti20nm/Pt20nm/100Au MS~2 at 3 different temperatures. Each measurement is the average of up to three samples (in some cases where the wires bonds broke, only 1 or 2 measurements were possible). The errorbars on the pull force are the standard deviation of the mean.  The vertical dashed line indicates the failure threshold for each data set based on pull strength. The grey box is the uncertainty in the time to failure as determined by the difference between the values for the fit (red solid line) of the pull force and the raw data. In cases where the sample did not fail (bottom, center), we use a linear fit to the resistance, blue dashed line, to extrapolate an anticipated time to failue. }
    \label{fig:thinGoldvsTemp}
\end{figure*}

Functional wire bonds must meet the criteria of low resistance and high pull strength to be used on the ion trap. 
The limit for wire bond resistance increase is set to 50~m$\Omega$ based on the results from the bond failure of the base condition of gold wire bonded to aluminum pad (see Fig.~\ref{fig:thinGoldvsTemp}). There is a sharp increase in resistance around the +50~m$\Omega$ point, where the resistance begins to rise much faster with time. Too much resistance on the rf electrode can contribute to heating and cause the IMC formation to run-away as up to 300~V of rf is applied. Also around the time the resistance increases by 50~m$\Omega$, the pull strength drops below 3~gf which is the industry standard for a bond failure as specified in MIL-STD-883 Method 2011 \cite{NASAMIL}.
It is shown in the results below that adding a metal barrier on top of the aluminum pad reduces the effects of IMC growth.
However, due to trade-off in functionality for thicker metal stacks we limit the maximum metal thickness to under $0.5~\mu$m, though there are cases where that limit is lower or higher depending on the end use and trap design.

\begin{table}[ht] 
    \begin{center}
    \caption{Time to failure for the tested metal stacks at different temperatures. Testing requires long times scales which limited the number of stacks tested at each temperature. 
    }
    \label{tab:Failure_values}    
    \begin{tabular}{|c|c|c|c|c|}
    \hline
            & Metal Stacks      & {\multirow{2}{*}{ 200$^{\circ}$C }}  &  {\multirow{2}{*}{ 300$^{\circ}$C }}  &   {\multirow{2}{*}{ 350$^{\circ}$C }}  \\
            & Ti/Pt/Au (nm)     &                   & &          \\ \specialrule{.15em}{.0em}{.0em} 
    MS 1    & 0/0/0             &   384$\pm$58 hrs     &    29$\pm$6 hrs      &   1.5$\pm$0.5 hrs    \\ \hline
    MS 2    & 20/20/50          &   672$\pm$24 hrs     & 46 hrs$^{*}$      &   2.5$\pm$0.5 hrs    \\ \hline
    MS 3    & 20/100/250        &   2100$\pm$400 hrs$^{\dagger}$   &   163 hrs$^{*}$       &   22 hrs$^{*}$         \\ \hline
    MS 4    & 20/100/1000       &   3264$\pm$318 hrs      &   240 hrs$^{\dagger}$           &   12 hrs$^{\dagger}$             \\ \hline
    MS 5    & 20/300/1000       &   60000 hrs$^{*}$ &   4500 hrs$^{\dagger}$&   230 hrs$^{\dagger}$\\ \hline
    \end{tabular}
    \end{center}
    \footnotesize{$^{*}$ Values determined from fitting the resistance as shown in Figs.~\ref{fig:thinGoldvsTemp} and ~\ref{fig:Hybrid_vs_temp}.}\\
    \footnotesize{$\dagger$ Failure extrapolated from other temperature data, shown in Fig.~\ref{fig:extrapolation} for MS-3}. 
\end{table}

\begin{figure*}[h]
    \centering
    \includegraphics[width=0.7\textwidth]{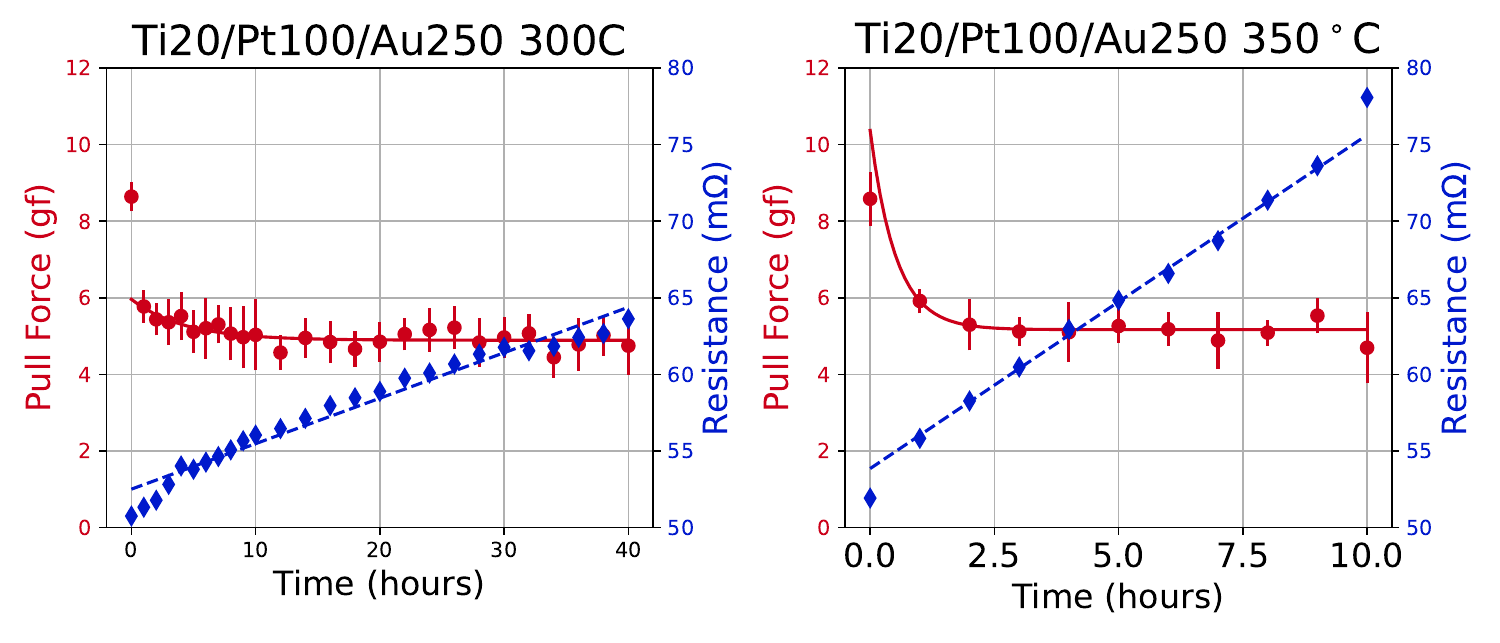}
    \caption{Measured pull force to break the wire bond and the resistance over time for a wedge bonded gold wire on MS~3 at 2 different temperatures. Each measurement is the average of up to three samples (in some cases where the wires bonds broke, only 1 or 2 measurements were possible). The errorbars on the pull force are the standard deviation of the mean.  Neither dataset crosses a threshold for failure throughout the measurement, though in each case failing from resistance is likely to occur first as the pull strength appears to have plateaued. From the results of the other metal stacks versus temperature, we can infer the success of this metal stack at 200$^{\circ}$C. }
    \label{fig:Hybrid_vs_temp}
\end{figure*}

As each of the samples progress through the aging bake tests the resistance, pull strengths, and pictures are recorded. In each of the bond pad metal stack combinations (listed in Table~\ref{tab:Failure_values}), the initial pull strength is relatively high and then the breaking force drops rapidly before it tapers off to a more gradual decline as the bake time increases. This initial drop in pull strength is attributed to the annealing of the gold wire. When the wire is annealed it reduces the breaking load of the wire. In the case of the bare aluminum pad, the pull strength continues to decline until the bonds lift due to IMC growth (see Fig.~\ref{fig:thinGoldvsTemp}).
However, the gold coated bond pads' pull strength fall in value slower, depending on the thickness of the gold and platinum (see Fig.~\ref{fig:thinGoldvsTemp} and Fig.~\ref{fig:Hybrid_vs_temp}). 
Historically, our traps had bare aluminum bond pads, which based on previous studies, limited our vacuum bakes to less than 7 days \cite{Maunz2016}. 
The results from those previous studies differ from the results in Table~\ref{tab:Failure_values} due to the geometry of the test structure and the exact wire bond welding parameters of the bonds.
Since a single bake can take 4 to 5 days, each ion trap is limited to a single use. 
Metal stacks that can survive more than 21 days would allow for 4-5 vacuum bakes and enable long term use of ion traps. 
Either MS~2 or MS~3 are suitable for application under this criteria alone. 

\begin{figure}
    \centering
    \includegraphics[width= 0.9\columnwidth]{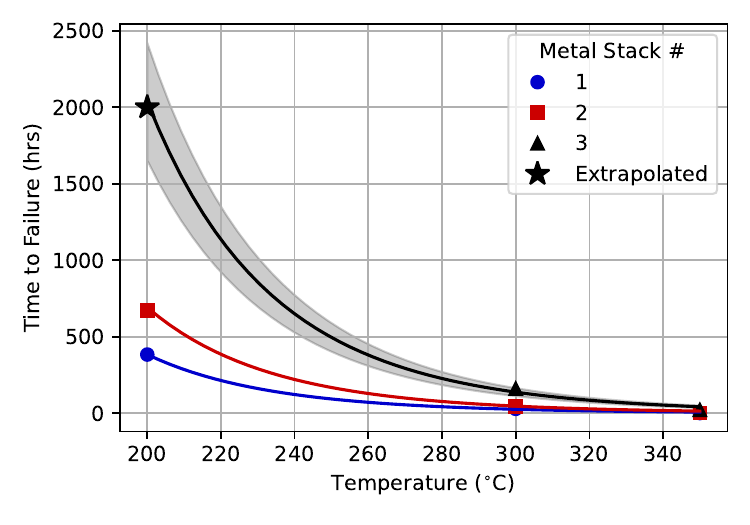}
    \caption{Time to failure vs temperature for MS-1 though MS-3. The data set for MS-1 is complete and fits reasonably to a phenomenological exponential fit $Ae^{-0.846\sqrt{T}}$. The data for MS-2 was fit to the same model allowing only the amplitude A to scale. Given the agreement of the fits, we used that model to extrapolate the 200$^{\circ}$C failure point for MS-3 (black star). The gray shaded region is the uncertainty given from scaling the amplitude to lower and higher limits of hitting the 300$^{\circ}$C point. }
    \label{fig:extrapolation}
\end{figure}

While pull strengths are important to make sure the bonds will stay attached, resistances of the wire bonds are the most important factors for ion traps as they directly affect the performance of the ion trap device. This is most notable in the case of gold wire on aluminum pads.
Initially, the resistance contribution per bond is small and therefore 4-point measurements are used on sections of several wires in series (see Fig.~\ref{fig:TestStructure}). 
In the case of the aluminum bond pads (MS~1), the bonds failed due to the resistance increasing more than 50~m$\Omega$ during the bake simultaneously to the pull strength dropping below 3~gf.
Most of the other samples show a rising resistance and in many cases the resistance will fail at nearly the same time or shortly after the pull strength. 
We do notice the trend that the initial rise is resistance is roughly linear, especially when the resistance is less than about 150~m$\Omega$, even for most of the aluminum samples. 
Thus, we use this as an indicator of when the metal stack will fail if it did not occur within our testing window, as shown in Fig.~\ref{fig:thinGoldvsTemp} and Fig.~\ref{fig:Hybrid_vs_temp}.
For cases where the MS was not tested in that temperature range, such as MS-3 at 200$^{\circ}$C, we use the scaling of failure time versus temperature of the other metal stacks to extrapolate the failure, as shown in Fig.~\ref{fig:extrapolation}.
Other work has shown that the failure versus temperature follows a general exponential trend~\cite{Ahmad2019}. 
Here, we use a phenomenological curve to fit MS-1 and MS-2 which allows us to infer a value for MS-3 at 200$^{\circ}$C. 
Only the amplitude of the fit was allowed to vary between the 3 data sets. 
Though this extrapolation is not based on many points, the failure on any particular wirebond can vary greatly from one bond to the next. 
We use these failure times (even the measured ones) as guidelines for performance and aim not to exceed at least 70\% of that time in a bake.
Additionally, the application of the wirebonds onto our traps, with the ultra-low profile will likely result in much shorter times as we see our test structures here last about twice as long for bare aluminum as compared to our previous measurements in \cite{Maunz2016}.

\begin{figure}[ht]
    \centering
    \subfloat[]{\includegraphics[width=0.95\columnwidth]{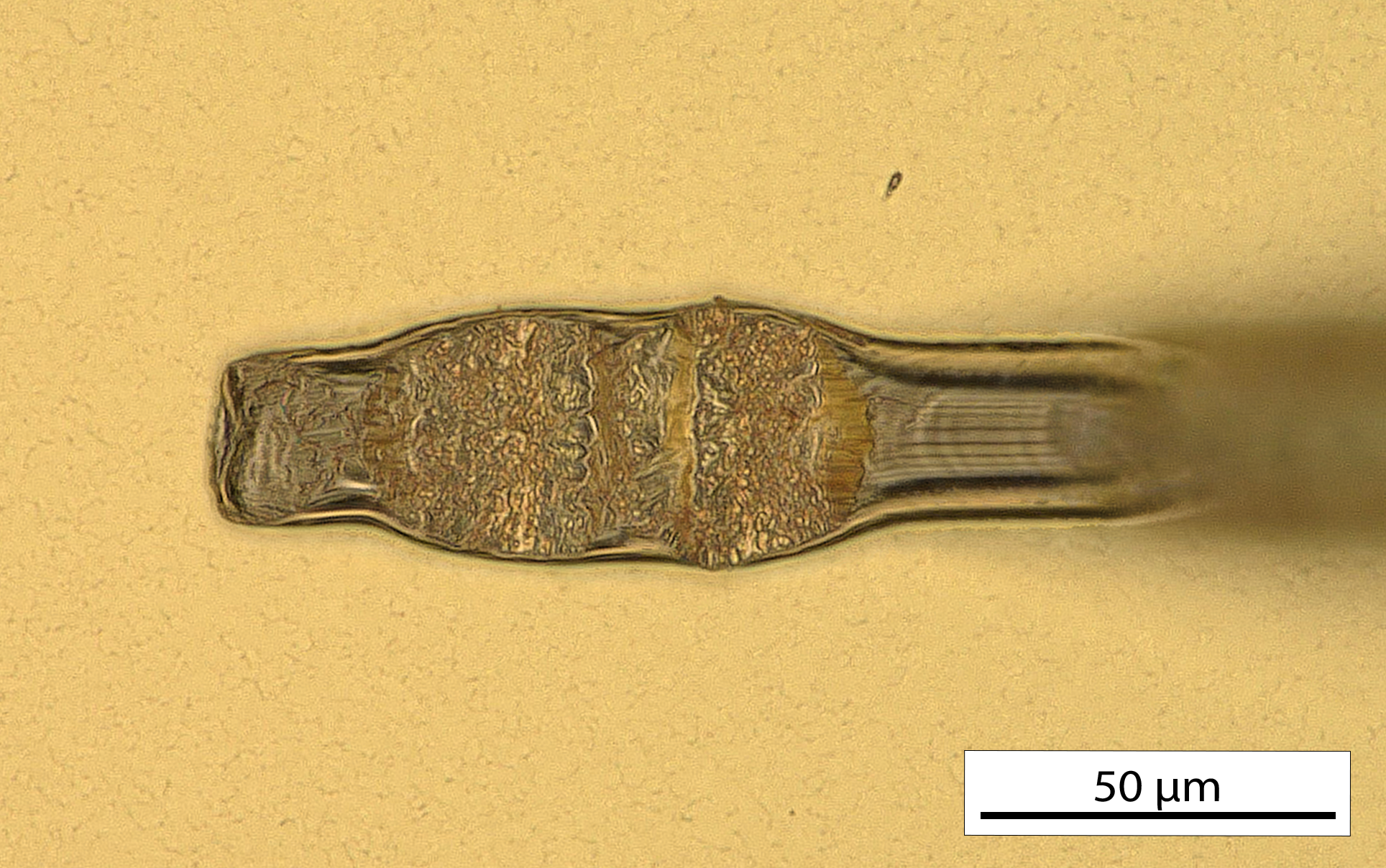}}
    \hfil
    \subfloat[]{\includegraphics[width=0.95\columnwidth]{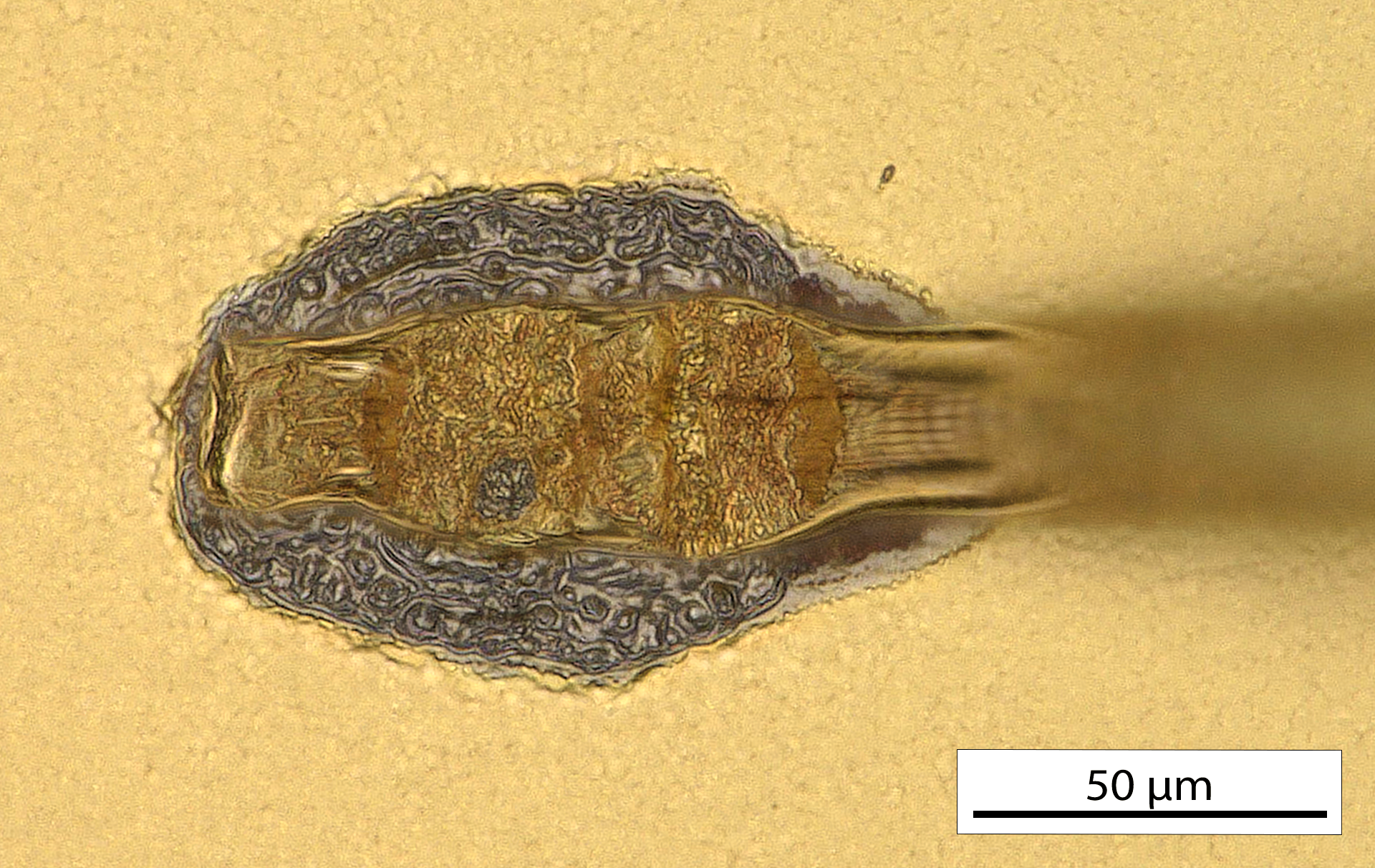}}
    \caption{Optical image of (a) before and (b) after bake for MS-3. Though significant amount of IMC is forming around the bond, the bond itself remains strong and low resistance.  }
    \label{fig:beforeafterbake}
\end{figure}

High-magnification optical (Fig.~\ref{fig:beforeafterbake}) and SEM images (Fig.~\ref{fig:SEM_no_bake_section} and Fig.~\ref{fig:SEM_29at200_section}) show the IMC growth outward from the bond.
In the case of bond pad metals that passed the pull test and resistance criteria, this outward growth is only detrimental if the bond pad pitch is small enough for the IMC to come in contact with adjacent pads causing electrical shorts. FIB cross-sections are useful to inspect the evolution of the weld during the high-temperature bakes. These FIB cross-sections (Fig.~\ref{fig:SEM_no_bake_section} and Fig.~\ref{fig:SEM_29at200_section}) show Kirkendall voids \cite{JI2007,Xie2021} and areas where there is a clear delineation between the wire and the pad. 

A summary of the results is shown in Table~\ref{tab:Failure_values}. 
Both of the thick gold metal stacks (MS~4 and MS~5) passed the resistance and pull tests at 200$^{\circ}$C and we can extrapolate the success of MS~3 at 200$^{\circ}$C based on passing the tests at higher temperatures.
For cases where additional metal will not impact device performance, thicker metal stacks are ideal. 
These stacks, MS~4 and MS~5, show minimal change in performance and negligible growth in the IMC on the surface of the bond pad after the 200$^{\circ}$C bake. 
Though these stacks may not be suitable for our ion traps, we include them to demonstrate that the thick barrier can overcome the IMC growth.
Additionally, we note that not all metal stacks were tested at all temperatures. 
This is due to the fact that some of the testing can take extreme amounts of time and when the original long tests were performed at 200$^{\circ}$C, it was not clear that we would want to test MS-3.

While we observed a general trend of thicker metal limiting IMC growth, we note that a thick platinum barrier may be the most important layer.  Indeed, a contributing factor to the success of MS~3 is likely the thick diffusion barrier offered by the platinum.
We did not test a variant of MS~2 with thick platinum, but if thinner stacks are required, using less gold is likely preferred. A variation of MS~3 was attempted with thick titanium, however, this metal stack proved difficult to wire bond and thus we were unable to compare it due to wire bonds failing during assembly.

Finally, optimizing the welding parameters can improve Kirkendall voiding and bond heel strength \cite{Schneider2009}. 
The bond parameters in this work were optimized for pull strength from the best known parameter space for the target device geometries. 
However, the time to failure could possibly be extended further by optimizing for bond shape or lowest resistance instead of pull strength, for example.
We choose to optimize the bonds for pull-strength as it has a quantitative comparison and simple implementation.
Additionally, we want each bond to be comparable across the metal stacks, so we use the same optimization protocol for each stack.
Further investigation could be done and possibly achieve even longer time to failure by optimizing for a different parameter such as final resistance after temperature cycling. 

We demonstrate that using a metal stack of Ti/Pt/Au as opposed to bare aluminum can almost double the bakable lifetime of the ion trap device, from 384 hours to 672 hrs for MS~2.
Based on extrapolating from performance at higher temperatures, we believe the failure of our preferred stack MS~3 may be approximately 2100 hours. 
This metal stack thus allows for multiple vacuum bakes without risking failure on a single ion trap device. 
Additionally, we suggest that thin titanium and thick platinum are key to successful wire bonding and mitigating IMC formation for cases where this metal stack is not applicable. 
While MS~3 does not prevent the formation of IMC (see fig.~\ref{fig:beforeafterbake}), the bond to this metal stack passes all criteria to be a good wire bond in this use case.
This solution is especially suited for applications where the quality of the metal surface is critical or a minimal thickness it required, such as ion trapping.

\section{Conclusion}

Gold-aluminum intermetallic growth at wire bonds causes increased brittleness and electrical resistance, which can lead to degradation or failure of ion trap devices.  
To prevent this failure mechanism, metal stacks can be applied to the aluminum surface to act as a diffusion barrier between the gold wire bond and the aluminum.  We found, not unsurprisingly, that thicker metal stacks led to better performance in terms of preventing intermetallic growth, however, thick stacks are also more likely to contribute to catastrophic failure and poor performance of the ion trap due to rf breakdown occurring at lower voltages.  
 
In this work, using accelerated aging tests and FIB cross-sections, we investigated adding thin layers of Ti/Pt/Au to aluminum pads.  We showed that a metal stack of Ti20nm/Pt100nm/Au250nm on a AlCu1.2$\mu$m bondpad was sufficient in slowing down the intermetallic growth to exceed our threshold of 21 days by lasting an estimated 2100hrs (86 days) at 200$^{\circ}$C.
In this time period, a single ion trap device could survive several vacuum bake cycles without risking the ion trap performance due to wire bond failure.

The ion traps discussed in this paper are complicated and time consuming to fabricate and prepare. Thus, increasing the functional lifetime of ion trap assemblies by applying a thin metal stack is a huge boon to the field of ion trap quantum computing using surface traps.  These results will enhance those experiments by removing the need to discard the ion trap every time the vacuum chamber is opened. 
This may also increase the operational runtime of the chips as heat is dissipated into the trap from the high voltage on the rf which can accelerate aging. 

This method of intermetallic growth mitigation can be used in any system that undergoes high-temperature processing during device packaging, lid seal and assembly, or system prep for things like UHV applications and is especially suited for applications where a thin gold coating will enhance the device performance. 

\section*{Acknowledgments}
We thank Tipp Jennings, Drew Hollowell, Jordan Gutierrez and Henry Dallo for their help in developing the samples and coating approach. Addionally, we thank Matt Blain and Ashlyn Burch for their helpful conversations.

\bibliographystyle{IEEEtran}
\bibliography{PPbib}

\begin{thebibliography}{10}
\providecommand{\url}[1]{#1}
\csname url@samestyle\endcsname
\providecommand{\newblock}{\relax}
\providecommand{\bibinfo}[2]{#2}
\providecommand{\BIBentrySTDinterwordspacing}{\spaceskip=0pt\relax}
\providecommand{\BIBentryALTinterwordstretchfactor}{4}
\providecommand{\BIBentryALTinterwordspacing}{\spaceskip=\fontdimen2\font plus
\BIBentryALTinterwordstretchfactor\fontdimen3\font minus
  \fontdimen4\font\relax}
\providecommand{\BIBforeignlanguage}[2]{{%
\expandafter\ifx\csname l@#1\endcsname\relax
\typeout{** WARNING: IEEEtran.bst: No hyphenation pattern has been}%
\typeout{** loaded for the language `#1'. Using the pattern for}%
\typeout{** the default language instead.}%
\else
\language=\csname l@#1\endcsname
\fi
#2}}
\providecommand{\BIBdecl}{\relax}
\BIBdecl

\bibitem{Cho2015}
D.-I. Cho, S.~Hong, M.~Lee, and T.~Kim, ``A review of silicon microfabricated
  ion traps for quantum information processing,'' \emph{Micro and Nano Systems
  Letters}, vol.~3, no.~2, 2015.

\bibitem{Tabakov2015}
\BIBentryALTinterwordspacing
B.~Tabakov, F.~Benito, M.~Blain, C.~R. Clark, S.~Clark, R.~A. Haltli, P.~Maunz,
  J.~D. Sterk, C.~Tigges, and D.~Stick, ``Assembling a ring-shaped crystal in a
  microfabricated surface ion trap,'' \emph{Phys. Rev. Appl.}, vol.~4, p.
  031001, Sep 2015. [Online]. Available:
  \url{https://link.aps.org/doi/10.1103/PhysRevApplied.4.031001}
\BIBentrySTDinterwordspacing

\bibitem{Maunz2016}
P.~Maunz, ``High optical access trap 2.0,'' Sandia National Laboratories,
  Albuquerque, NM, Tech. Rep. SAND2016-0796R, 2016.

\bibitem{Revelle2020}
M.~C. Revelle, ``Phoenix and peregrine ion traps,'' 2020.

\bibitem{Blain2021}
M.~G. Blain, R.~Haltli, P.~Maunz, C.~D. Nordquist, M.~Revelle, and D.~Stick,
  ``Hybrid {MEMS}-{CMOS} ion traps for {NISQ} computing,'' \emph{Quantum Sci.
  and Technol.}, vol.~6, no.~3, p. 034011, jun 2021.

\bibitem{Moses2023}
S.~A. Moses, C.~H. Baldwin, M.~S. Allman, R.~Ancona, L.~Ascarrunz, C.~Barnes,
  J.~Bartolotta, B.~Bjork, P.~Blanchard, M.~Bohn, J.~G. Bohnet, N.~C. Brown,
  N.~Q. Burdick, W.~C. Burton, S.~L. Campbell, J.~P. C.~I. au2, C.~Carron,
  J.~Chambers, J.~W. Chan, Y.~H. Chen, A.~Chernoguzov, E.~Chertkov, J.~Colina,
  J.~P. Curtis, R.~Daniel, M.~DeCross, D.~Deen, C.~Delaney, J.~M. Dreiling,
  C.~T. Ertsgaard, J.~Esposito, B.~Estey, M.~Fabrikant, C.~Figgatt, C.~Foltz,
  M.~Foss-Feig, D.~Francois, J.~P. Gaebler, T.~M. Gatterman, C.~N. Gilbreth,
  J.~Giles, E.~Glynn, A.~Hall, A.~M. Hankin, A.~Hansen, D.~Hayes, B.~Higashi,
  I.~M. Hoffman, B.~Horning, J.~J. Hout, R.~Jacobs, J.~Johansen, L.~Jones,
  J.~Karcz, T.~Klein, P.~Lauria, P.~Lee, D.~Liefer, C.~Lytle, S.~T. Lu,
  D.~Lucchetti, A.~Malm, M.~Matheny, B.~Mathewson, K.~Mayer, D.~B. Miller,
  M.~Mills, B.~Neyenhuis, L.~Nugent, S.~Olson, J.~Parks, G.~N. Price, Z.~Price,
  M.~Pugh, A.~Ransford, A.~P. Reed, C.~Roman, M.~Rowe, C.~Ryan-Anderson,
  S.~Sanders, J.~Sedlacek, P.~Shevchuk, P.~Siegfried, T.~Skripka, B.~Spaun,
  R.~T. Sprenkle, R.~P. Stutz, M.~Swallows, R.~I. Tobey, A.~Tran, T.~Tran,
  E.~Vogt, C.~Volin, J.~Walker, A.~M. Zolot, and J.~M. Pino, ``A race track
  trapped-ion quantum processor,'' 2023.

\bibitem{Monroe2013}
\BIBentryALTinterwordspacing
C.~Monroe and J.~Kim, ``Scaling the ion trap quantum processor,''
  \emph{Science}, vol. 339, no. 6124, pp. 1164--1169, 2013. [Online].
  Available: \url{https://www.science.org/doi/abs/10.1126/science.1231298}
\BIBentrySTDinterwordspacing

\bibitem{Lacroute2016}
\BIBentryALTinterwordspacing
C.~Lacroûte, M.~Souidi, P.-Y. Bourgeois, J.~Millo, K.~Saleh, E.~Bigler,
  R.~Boudot, V.~Giordano, and Y.~Kersalé, ``Compact yb+ optical atomic clock
  project: design principle and current status,'' \emph{Journal of Physics:
  Conference Series}, vol. 723, no.~1, p. 012025, jun 2016. [Online].
  Available: \url{https://dx.doi.org/10.1088/1742-6596/723/1/012025}
\BIBentrySTDinterwordspacing

\bibitem{Ivory2021}
\BIBentryALTinterwordspacing
M.~Ivory, W.~J. Setzer, N.~Karl, H.~McGuinness, C.~DeRose, M.~Blain, D.~Stick,
  M.~Gehl, and L.~P. Parazzoli, ``Integrated optical addressing of a trapped
  ytterbium ion,'' \emph{Phys. Rev. X}, vol.~11, p. 041033, Nov 2021. [Online].
  Available: \url{https://link.aps.org/doi/10.1103/PhysRevX.11.041033}
\BIBentrySTDinterwordspacing

\bibitem{House2008}
\BIBentryALTinterwordspacing
M.~G. House, ``Analytic model for electrostatic fields in surface-electrode ion
  traps,'' \emph{Phys. Rev. A}, vol.~78, p. 033402, Sep 2008. [Online].
  Available: \url{https://link.aps.org/doi/10.1103/PhysRevA.78.033402}
\BIBentrySTDinterwordspacing

\bibitem{Wesenberg2008}
\BIBentryALTinterwordspacing
J.~H. Wesenberg, ``Electrostatics of surface-electrode ion traps,'' \emph{Phys.
  Rev. A}, vol.~78, p. 063410, Dec 2008. [Online]. Available:
  \url{https://link.aps.org/doi/10.1103/PhysRevA.78.063410}
\BIBentrySTDinterwordspacing

\bibitem{Wilson2022}
\BIBentryALTinterwordspacing
J.~M. Wilson, J.~N. Tilles, R.~A. Haltli, E.~Ou, M.~G. Blain, S.~M. Clark, and
  M.~C. Revelle, ``{In situ detection of RF breakdown on microfabricated
  surface ion traps},'' \emph{Journal of Applied Physics}, vol. 131, no.~13, p.
  134401, 04 2022. [Online]. Available: \url{https://doi.org/10.1063/5.0082740}
\BIBentrySTDinterwordspacing

\bibitem{Longo1963}
T.~A. Longo and B.~Selikson, ``Identification and elimination of a failure
  mechanism in semiconductor devices,'' in \emph{Second Annual Symposium on the
  Physics of Failure in Electronics}, 1963, pp. 424--435.

\bibitem{Selikson1964}
B.~Selikson and T.~Longo, ``A study of purple plague and its role in integrated
  circuits,'' \emph{Proceedings of the IEEE}, vol.~52, no.~12, pp. 1638--1641,
  1964.

\bibitem{Chen1967}
G.~Chen, ``On the physics of purple-plague formation, and the observation of
  purple plague in ultrasonically-joined gold-aluminum bonds,'' \emph{IEEE
  Transactions on Parts, Materials and Packaging}, vol.~3, no.~4, pp. 149--153,
  1967.

\bibitem{Philofsky1970}
\BIBentryALTinterwordspacing
E.~Philofsky, ``Intermetallic formation in gold-aluminum systems,''
  \emph{Solid-State Electronics}, vol.~13, no.~10, pp. 1391--1394, 1970.
  [Online]. Available:
  \url{https://www.sciencedirect.com/science/article/pii/0038110170901723}
\BIBentrySTDinterwordspacing

\bibitem{Philofsky1970_2}
E.~Philofski, ``Purple plague revisited,'' in \emph{8th Reliability Physics
  Symposium}, 1970, pp. 177--185.

\bibitem{Horsting1972}
C.~W. Horsting, ``Purple plague and gold purity,'' in \emph{10th Reliability
  Physics Symposium}, 1972, pp. 155--158.

\bibitem{Ahmad2019}
S.~S. Ahmad and S.~C. Smith, ``Au/al wire bond interface resistance degradation
  rate simulations,'' \emph{IEEE Transactions on Device and Materials
  Reliability}, vol.~19, no.~4, pp. 774--781, 2019.

\bibitem{Blish2007}
R.~C. Blish, S.~Li, H.~Kinoshita, S.~Morgan, and A.~F. Myers, ``Gold–aluminum
  intermetallic formation kinetics,'' \emph{IEEE Transactions on Device and
  Materials Reliability}, vol.~7, no.~1, pp. 51--63, 2007.

\bibitem{JI2007}
\BIBentryALTinterwordspacing
H.~Ji, M.~Li, C.~Wang, H.~S. Bang, and H.~S. Bang, ``Comparison of interface
  evolution of ultrasonic aluminum and gold wire wedge bonds during thermal
  aging,'' \emph{Materials Science and Engineering: A}, vol. 447, no.~1, pp.
  111--118, 2007. [Online]. Available:
  \url{https://www.sciencedirect.com/science/article/pii/S0921509306022921}
\BIBentrySTDinterwordspacing

\bibitem{Shepherd2010}
D.~Shepherd, P.~Grant, C.~Johnston, and S.~Riches, ``The behaviour of au-au
  wire bonds in extreme environments,'' \emph{IMAPSource Proceedings}, vol.
  2010, no. HITEC, pp. 114--121, 2010.

\bibitem{Harman2010}
\BIBentryALTinterwordspacing
G.~Harman, \emph{\BIBforeignlanguage{en}{Wire Bonding in Microelectronics}},
  3rd~ed.\hskip 1em plus 0.5em minus 0.4em\relax New York: McGraw-Hill
  Education, 2010. [Online]. Available:
  \url{https://www.accessengineeringlibrary.com/content/book/9780071476232}
\BIBentrySTDinterwordspacing

\bibitem{DeLucca2012}
\BIBentryALTinterwordspacing
J.~DeLucca, J.~Osenbach, and F.~Baiocchi, ``Observations of imc formation for
  au wire bonds to al pads,'' \emph{Journal of Electronic Materials}, vol.~41,
  no.~4, pp. 748--756, Apr 2012. [Online]. Available:
  \url{https://doi.org/10.1007/s11664-011-1805-8}
\BIBentrySTDinterwordspacing

\bibitem{Kim2013}
H.~J. Kim, M.-S. Song, K.-W. Paik, J.-T. Moon, and J.-Y. Song, ``Investigation
  of interfacial phenomena of alloyed au wire bonding,'' in \emph{2013 IEEE
  15th Electronics Packaging Technology Conference (EPTC 2013)}, 2013, pp.
  479--482.

\bibitem{Xie2021}
S.~Xie, P.~Lin, and Q.~Yao, ``Overview of au-al bond interface,'' in \emph{2021
  International Conference on Electronics, Circuits and Information Engineering
  (ECIE)}, 2021, pp. 250--253.

\bibitem{Seigal1996}
\BIBentryALTinterwordspacing
P.~Seigal, R.~Briggs, D.~Rieger, A.~Baca, and A.~Howard, ``Adhesion studies of
  gaas-based ohmic contact and bond pad metallization,'' \emph{Thin Solid
  Films}, vol. 290-291, pp. 503--507, 1996, papers presented at the 23rd
  International Conference on Metallurgical Coatings and Thin Films. [Online].
  Available:
  \url{https://www.sciencedirect.com/science/article/pii/S0040609096090219}
\BIBentrySTDinterwordspacing

\bibitem{Stareev1993}
\BIBentryALTinterwordspacing
G.~Stareev, H.~Künzel, and G.~Dortmann, ``{A controllable mechanism of forming
  extremely low‐resistance nonalloyed ohmic contacts to group III‐V
  compound semiconductors},'' \emph{Journal of Applied Physics}, vol.~74,
  no.~12, pp. 7344--7356, 12 1993. [Online]. Available:
  \url{https://doi.org/10.1063/1.355002}
\BIBentrySTDinterwordspacing

\bibitem{Schneider2009}
M.~Schneider-Ramelow, S.~Schmitz, B.~Schuch, and W.~Grubl, ``Kirkendall voiding
  in au ball bond interconnects on al chip metallization in the temperature
  range from 100 – 200°c after optimized intermetallic coverage,'' in
  \emph{2009 European Microelectronics and Packaging Conference}, 2009, pp.
  1--6.

\bibitem{NASAMIL}
\BIBentryALTinterwordspacing
``Test method standard - microcircuits.'' [Online]. Available:
  \url{https://s3vi.ndc.nasa.gov/ssri-kb/static/resources/std883.pdf}
\BIBentrySTDinterwordspacing

\end{thebibliography}

\newpage

\section{Biography Section}

\begin{IEEEbiography}[{\includegraphics[width=1in,height=1.25in,clip,keepaspectratio]{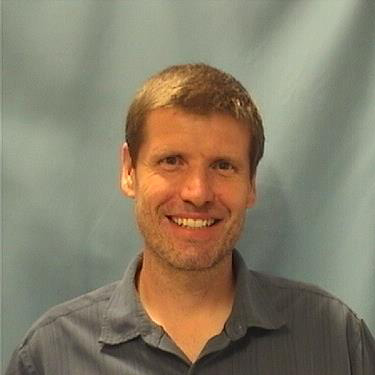}}]{Raymond A Haltli}
obtained B.S. degree in Electronics Engineering Technology from the DeVry University, Phoenix, Arizona in 2003. In 2015 he received a M.S. degree in Electrical Engineering from The University of New Mexico.

Currently he is a Senior Member of the Technical Staff at Sandia National Laboratories. He join Sandia in 2004 and is currently the technology lead for the assembly and packaging of silicon-based surface electrode ion traps for quantum information sciences and other silicon-based devices.  

\end{IEEEbiography}

\begin{IEEEbiography}[{\includegraphics[width=1in,height=1.25in,clip,keepaspectratio]{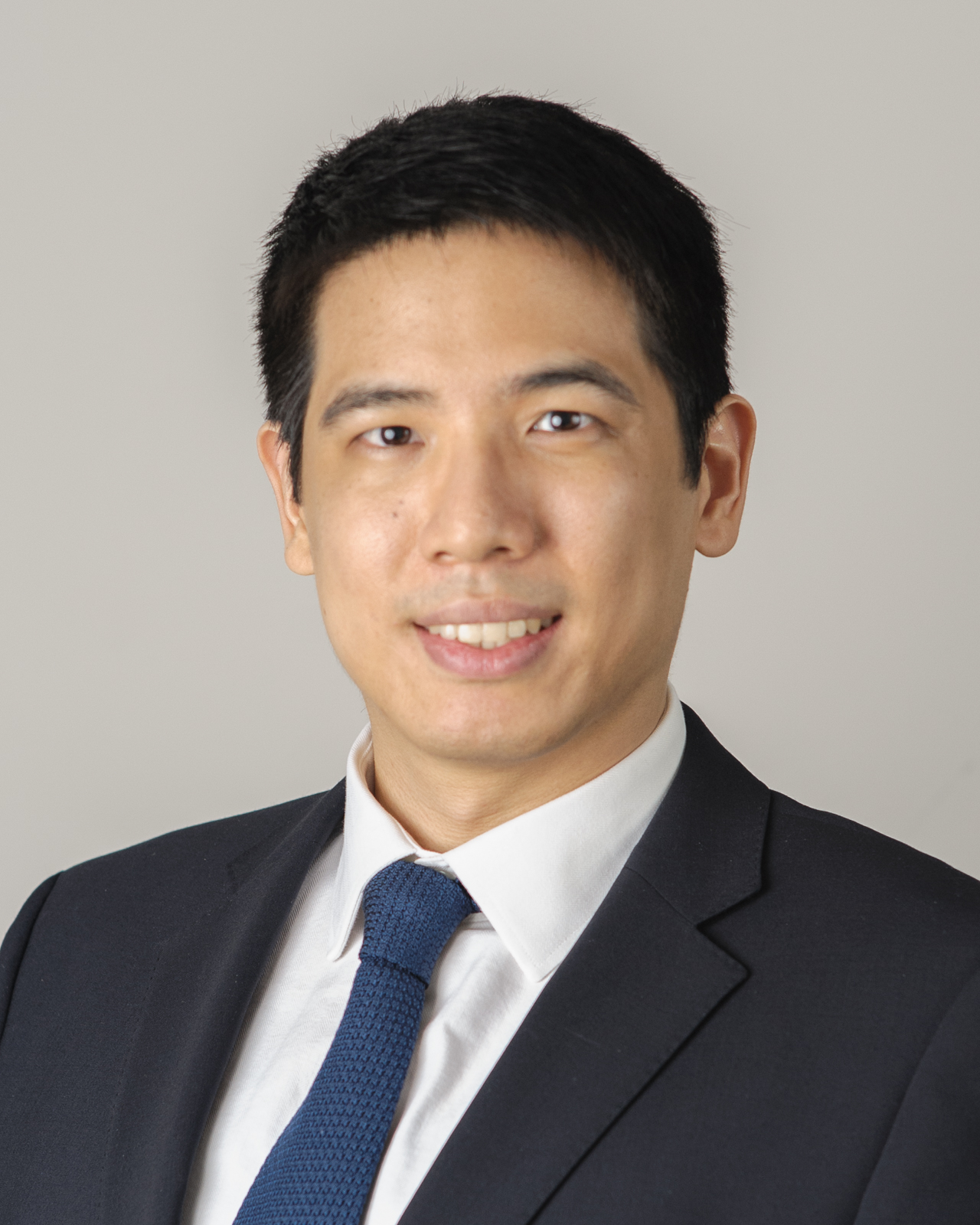}}]{Eric Ou}
obtained B.S., M.S., and Ph.D. degrees in Mechanical Engineering from the University of Texas at Austin in 2012, 2015, and 2020, respectively, with a concentration in thermal fluid systems and nanoscale heat transfer.

He joined Sandia National Laboratories in 2020 and currently works on ion trap integration and fabrication as a Senior Member of the Technical Staff.  

\end{IEEEbiography}

\begin{IEEEbiography}[{\includegraphics[width=1in,height=1.25in,clip,keepaspectratio]{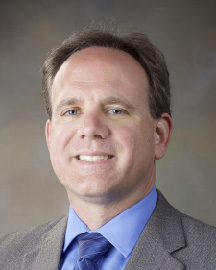}}]{Christopher D. Nordquist}
received B.S., M.S., and Ph. D. degrees in electrical engineering from The Pennsylvania State University in 1997, 1998, and 2002.  At Penn State he was a research assistant from 1995-1998 and NDSEG Fellow from 1998-2001, where he explored technology integration using self-assembly.
In 2002, he joined Sandia National Laboratories, where he is currently a Distinguished Member of Technical Staff in the RF Microsystems Department.  His research interests include the realization and integration of emerging micromachined and solid-state devices for high-speed and radio frequency applications.  Dr. Nordquist is a Senior Member of the IEEE and has co-authored over 90 publications and holds 12 patents.

\end{IEEEbiography}

\begin{IEEEbiography}[{\includegraphics[width=1in,height=1.25in,clip,keepaspectratio]{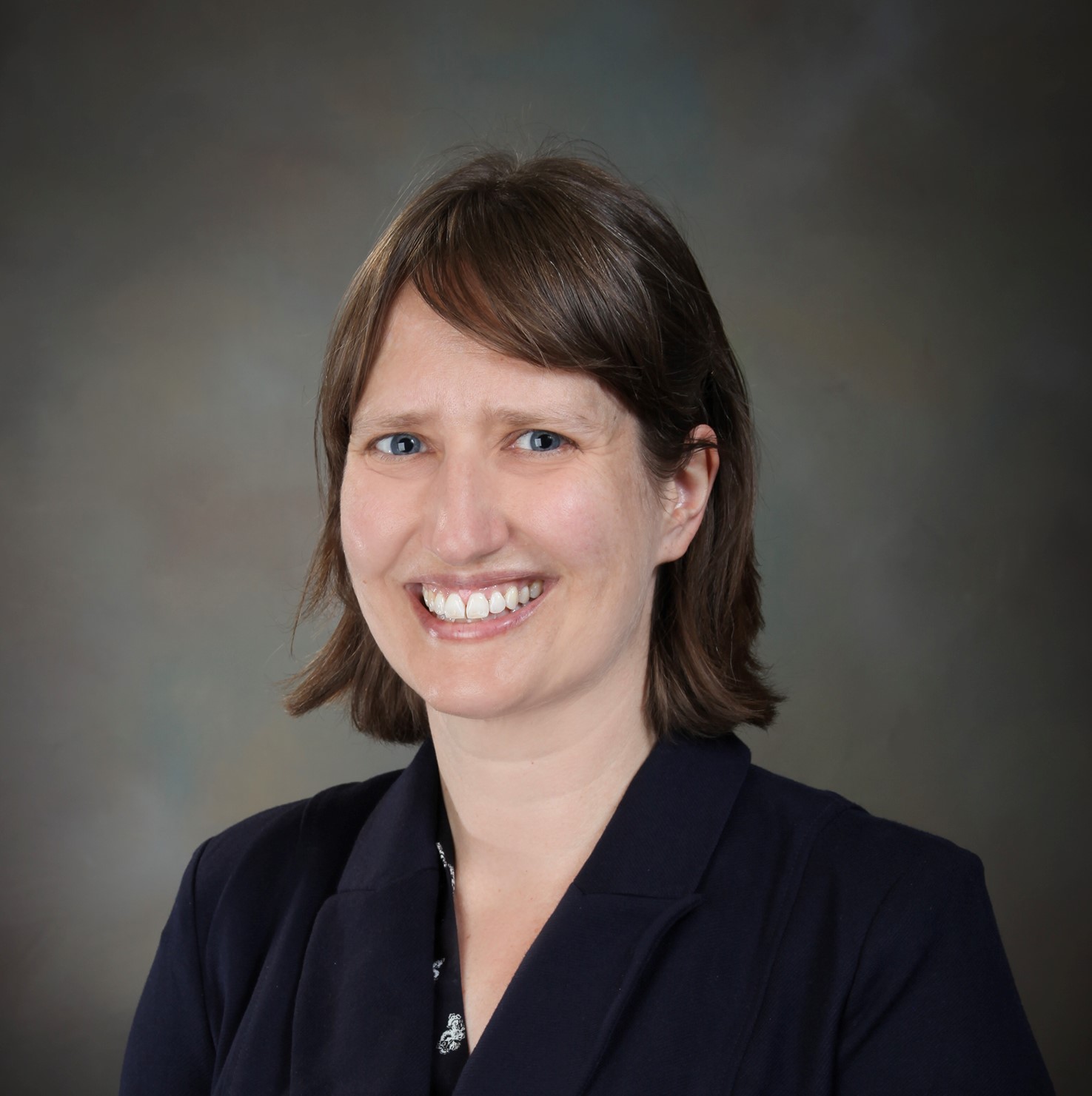}}]{Susan M. Clark} was born in Alexandria, Virgina in 1982. She received a B.S. degree in physics from Duke University, Durham, North Carolina in 2004.  In 2010, she received a Ph.D. degree in applied physics from Stanford University, Stanford, California for work on optical control of qubits in semiconductors.     

From 2010 to 2013, she was a Joint Quantum Institute Postdoctoral Fellow at the University of Maryland, College Park, Maryland.  At that time, she worked with Christopher Monroe demonstrating protocols for combining remote and local entanglement for scalable trapped-ion quantum systems.  She joined Sandia National Laboratories in 2013, and is currently a Distinguished Member of the Technical Staff.  Since joining Sandia, she has worked on a variety of quantum information related projects involving trapped ions and gate defined quantum dots in Silicon.  She is currently the PI of the DOE Office of Science ASCR funded program QSCOUT (Quantum Scientific Computing Open User Testbed).  
\end{IEEEbiography}

\begin{IEEEbiography}[{\includegraphics[width=1in,height=1.25in,clip,keepaspectratio]{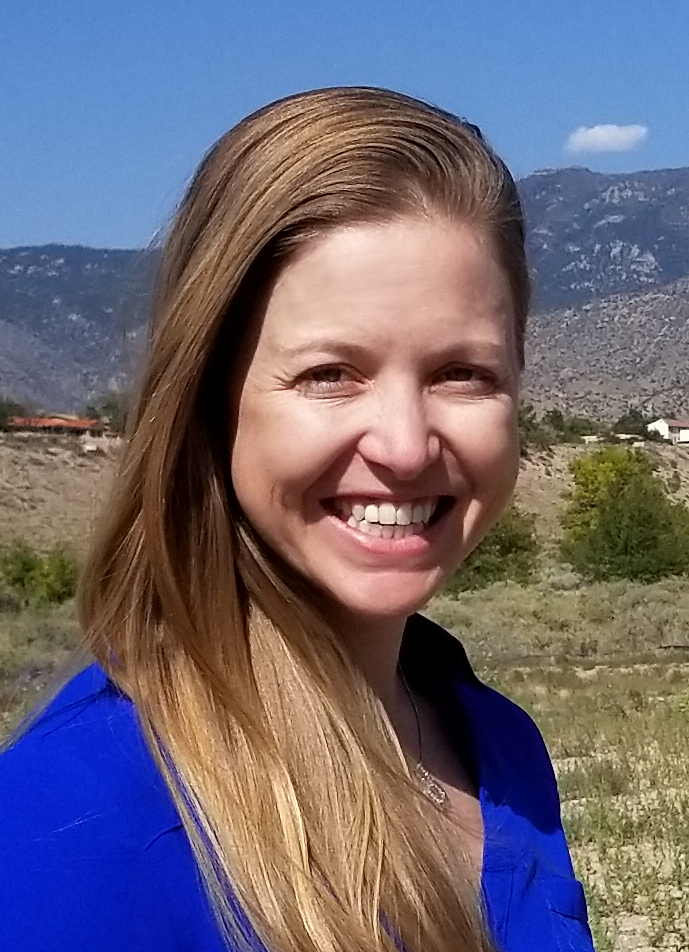}}]{Melissa C. Revelle} received B.S. degrees in Physics and Astronomy from the University of Arizona in 2009. She earned her M.S. degree in physics in 2013 and a Ph.D. in atomic physics in 2016, both from Rice University in Houston, Texas. 

Currently, she is a Principal Member of Technical Staff at Sandia National Laboratories in Albuquerque, New Mexico. She worked as a research assistant studying the polarizability of atoms from 2007-2009 at the University of Arizona and investigating the dimensional crossover in ultra-cold fermi gases from 2009-2016 at Rice University. Her research interests currently include coherent manipulation and control of trapped ions in microfabricated surface traps. 
Dr. Revelle’s awards include the Wilson Research award from Rice University in 2017. 

\end{IEEEbiography}

\vfill

\end{document}